\documentclass[twocolumn,showpacs,preprintnumbers,amsmath,amssymb]{revtex4}

\include{sgdef}
\usepackage{graphicx}
\usepackage{amsmath}
\usepackage{dcolumn}
\usepackage{bm}

\begin{document}

\preprint{APL/StripMKID}

\title{A Position Sensitive X-ray Spectrophotometer using Microwave Kinetic Inductance Detectors}

\author{Benjamin A. Mazin, Bruce Bumble, and Peter K. Day}

\affiliation{Jet Propulsion Laboratory, California Institute of
Technology}

\author{Megan E. Eckart, Sunil Golwala, Jonas Zmuidzinas, and Fiona A.
Harrison}

\affiliation{Physics Department, California Institute of Technology}

\date{\today}

\begin{abstract}

The surface impedance of a superconductor changes when energy is
absorbed and Cooper pairs are broken to produce single electron
(quasiparticle) excitations. This change may be sensitively measured
using a thin-film resonant circuit called a microwave kinetic
inductance detector (MKID). The practical application of MKIDs for
photon detection requires a method of efficiently coupling the
photon energy to the MKID. We present results on position sensitive
X-ray detectors made by using two aluminum MKIDs on either side of a
tantalum photon absorber strip. Diffusion constants, recombination
times, and energy resolution are reported. MKIDs can easily be
scaled into large arrays.

\end{abstract}



\maketitle

Low temperature detectors (LTDs) are the detectors of choice to
measure the energy and arrival time of incoming single photons.
Arrays of LTDs can determine the location, time, and energy of every
incoming photon (imaging spectrophotometry) with no read noise or
dark current. Many technologies are being developed, including
neutron transmutation doped (NTD) germanium\cite{astroe2},
superconducting tunnel junctions
(STJ)\cite{kraus86,friedrich97,li03}, transition edge sensors
(TES)\cite{irwin96,cabrera98,irwin00}, magnetic microcalorimeters
(MMC)\cite{Ens02}, and normal-insulator-superconductor (NIS)
bolometers\cite{schmidt05}. While these technologies have shown
promise in single pixel and small array devices, multiplexed
readouts remain a significant challenge and have only been
demonstrated for TES detectors, using complex superconducting
circuitry at 4 K (or colder)\cite{chervenak99,yoon01}.

The uses of energy-resolving X-ray detectors are both practical and
exotic.  High resolution X-ray detectors are used in X-ray
microanalysis to investigate semiconductor fabrication
problems\cite{wollman97,wollman00}, but could also be used to learn
about the strong gravitational fields around supermassive black
holes\cite{weaver99}. The work described here can also be adapted to
optical/UV energy-resolved single photon detection by increasing the
responsivity of the detectors. Imaging optical spectrophotometers
have a variety of astronomical applications, including planet
detection, optical pulsars\cite{romani01}, and redshift
determination of high-$z$ galaxies\cite{mazin00}.

An energy-resolving detector for photon energies of $0.1 \sim 10$
keV can be made using a ``strip-detector" architecture (Figure 1),
comprising a long strip of a superconducting material with
quasiparticle sensors attached at each end\cite{kraus86}. The
quasiparticle sensors we use are microwave kinetic inductance
detectors (MKIDs)\cite{day03}, and will be discussed in detail
below.  STJs have been previously used with this type of detector
architecture\cite{kraus86,friedrich97,li03}.

The photon detection process begins when an X-ray with energy $h\nu$
is absorbed in a tantalum strip, producing a number of excitations,
called quasiparticles, equal to $N_{qp} = \eta h \nu / \Delta$,
where $\Delta$ is the gap parameter of the superconductor, and
$\eta$ is an efficiency factor\cite{kozorezov00} (about 0.6 for our
devices). The principle is similar to electron-hole generation by
photons in semiconducting X-ray detectors, with the difference that
$\Delta$ is only tenths of millielectron volts (meV), as opposed to
1 eV or more for a semiconductor. This very low gap energy means
that millions of quasiparticle excitations are created for each
X-ray photon absorbed.  Since some of the energy is lost to phonons,
the fundamental energy resolution of the detector is limited by the
statistical fluctuation of the number of remaining quasiparticles,
given by $\sigma_N = \sqrt{F N_{qp}}$, where $F$ is the Fano
factor\cite{fano47}.

\begin{figure}
\begin{center}
\includegraphics[width=\columnwidth]{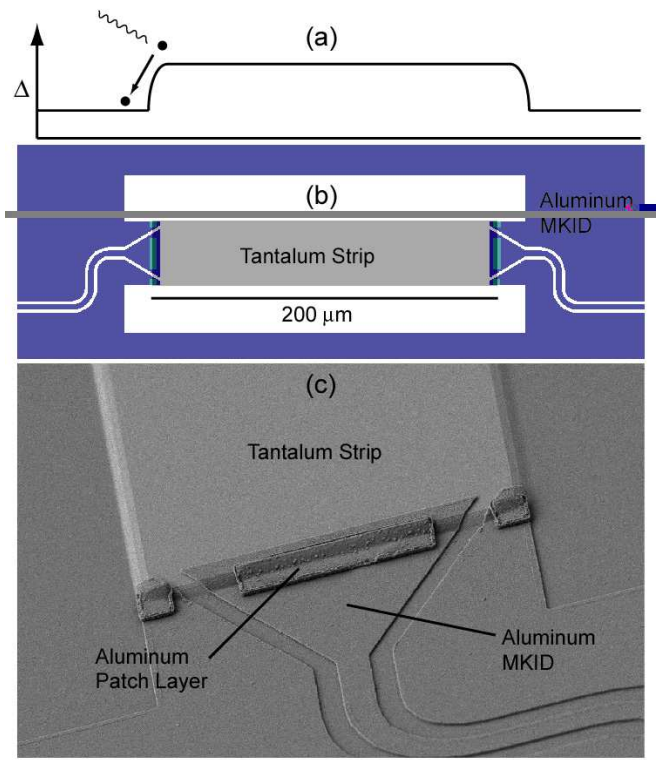}
\end{center}
\caption{The middle panel, Figure 1(b) contains a drawing of the
central region of a MKID strip detector. A $200 \times 35$~$\mu$m,
600 nm thick tantalum strip (RRR=22.6) is fabricated on R-plane
sapphire and has MKIDs attached to both ends. The 3~$\mu$m center
strip of the 200 nm thick aluminum (RRR=9.5) CPW resonator that
comprises the MKID is flared out where it contacts the tantalum
strip to allow lateral trapping of quasiparticles.  The top panel,
Figure 1(a) shows the superconducting gap $\Delta$ of the structure,
including a quasiparticle diffusing into the aluminum MKID and being
trapped by phonon emission.  The bottom panel, Figure 1(c) shows a
SEM of the Al-Ta interface from the wafer tested in this paper.   A
patch of aluminum patterned with a liftoff process is used to bridge
the Al-Ta interface to avert a step coverage problem.  In this
device the tantalum is nicely sloped and the aluminum resonator
climbs smoothly over the step.} \label{fig1}
\end{figure}

The tantalum absorber strip has a higher superconducting energy gap
($\Delta = 0.67$~meV) than the aluminum MKIDs ($\Delta = 0.18$~meV).
The quasiparticles created by photons absorbed in the tantalum strip
may diffuse laterally, reaching the MKIDs at the two ends of the
strip. Once in the MKIDs, the quasiparticles quickly cool by phonon
emission. This energy loss prevents them from returning to the
higher gap tantalum absorber, trapping them in the MKID. This
trapped quasiparticle population is measured by the MKIDs. The two
MKID output signals may be used to simultaneously deduce the
position and energy of the event. Noise sources will produce some
scatter $\delta E$ in the energy and $\delta x$ in position; the
fractional resolutions in energy and position are expected to be
comparable.

The quasiparticles trapped in the MKIDs are sensed through their
effect on the kinetic inductance and surface resistance of the
aluminum film comprising the MKID. The MKIDs are microwave
resonators made using coplanar waveguide (CPW) transmission
lines\cite{day03}. The 3~$\mu$m  wide CPW center strip is separated
from the ground plane by slots that are 2~$\mu$m wide. The length of
the resonator is $\sim5$ mm and the thickness of the Al film is 200
nm. An increase in the quasiparticle population in the MKID moves
the resonance frequency lower and increases the width of the
resonance (lower quality factor $Q$). Both of these effects are
monitored by measuring the amplitude and phase of a microwave probe
signal\cite{day03}.

The device shown in Figure 1 was cooled to 150 mK using an Oxford
Kelvinox 25 at the Caltech MKID test facility\cite{mazin04}. The
test sample contained eight separate strip detectors, with strip
lengths of 100, 200, 400, and 800~$\mu$m and strip widths of
35~$\mu$m, and four additional test resonators. For each strip
length, two different MKID designs were used, differing in the
strength of the coupling to a CPW readout line. The coupling
strength is specified by the corresponding coupling-limited quality
factor $Q_c$, which was chosen to be 25,000 and 50,000 for the two
MKID designs. All twenty MKID resonators were coupled to a single
feedline: the two resonators for a given strip were separated by 10
MHz in frequency, beginning at 6.5 GHz, while a 100 MHz spacing was
used to separate the different strip detectors.  All the resonators
were detected near their design frequencies.

Fabrication of this device is done on an R-plane sapphire wafer to
allow epitaxial growth of $\alpha$-phase (bcc) tantalum.  All metal
depositions are carried out in a load-locked ultra high vacuum (UHV)
sputtering system with a base pressure of $10^{-7}$~Pa.  The Ta film
is deposited at 60 nm/minute to a thickness of 600 nm with substrate
temperature of 700 C.  Our layers are patterned using a Canon 3000
stepping mask aligner with a Cymer 250 nm laser.  The tantalum film
is reactive ion etched (RIE).  Tantalum edge sloping is accomplished
by re-flowing the resist for 5 minutes at 130 C, followed by RIE
using a gas mixture of 30\% O$_2$ in CF$_4$ at a pressure of 27 Pa.
The resist is eroded back as the tantalum is removed. After the
surface is solvent cleaned, it is argon ion cleaned in-situ before
the aluminum for the MKID is blanket deposited to a thickness of 200
nm. RIE of aluminum is done with a mixture of 2:1, BCl$_3$ : Cl$_2$
at a pressure of 4 Pa. A water rinse to remove chlorine compounds is
followed by a solvent clean.

The device was illuminated with a weak $^{55}$Fe source that emits
Mn X-rays at $K_{\alpha}=5.9$ and $K_{\beta}=6.4$ keV.   In order to
collect X-ray data we first determine the resonant frequency of the
MKIDs from a frequency sweep.  Two microwave synthesizers are then
used to simultaneously excite and monitor the two MKIDs connected to
a given strip. X-rays absorbed in that strip produce large nearly
simultaneous pulses in the phases and amplitudes of the two
microwave readout signals. The rise times of the pulses are
controlled primarily by the diffusion time in the tantalum strip,
while the fall times are set by the quasiparticle lifetime in the
aluminum MKIDs. Each of the four readout channels (2 MKIDs, the
projection of amplitude and phase into rectangular coordinates for
each) is sampled at 250 kHz with 16-bit resolution and recorded.

After data collection we use an optimal filter to determine the
maximum pulse height in both channels.  This optimal filter is made
from a pulse template constructed by averaging many pulses that
occur near the center of the strip, and using the measured noise
spectrum from the MKID.  The simplified initial analysis presented
in this paper uses only the phase data; however, there is
significant information in the amplitude excursion, which we plan to
use in a later, more thorough analysis.  While the magnitude of the
amplitude signal is only about 20\% of the phase signal, the
amplitude noise, especially at low frequencies, can be significantly
lower.

This phase pulse data for the 200~$\mu$m strip is plotted for both
resonators in Figure 2.  We select ten $K_{\alpha}$ X-ray events
with absorption locations spread evenly over the absorber strip. The
detailed phase pulse shapes from these events are then used to
determine the relevant physical parameters of the device by fitting
to a diffusion-recombination model. While each X-ray event is
allowed to have a unique absorption location, all ten events are fit
using a single set of values for the diffusion constant and
quasiparticle lifetime in the tantalum strip and the quasiparticle
lifetimes in the two aluminum MKIDs. In addition, a scaling factor
accounting for the differing responsivities of the two MKIDs is
introduced by allowing a linear prefactor to modify the responsivity
of the left MKID.  The model also includes a recombination constant
(which is the same for both MKIDs) that depends on quasiparticle
density in the MKID.  This is used to model the enhanced
recombination in the aluminum MKID that can occur at the beginning
of a pulse if the quasiparticle density is high.

\begin{table}
\begin{tabular}{l c}
  \hline
  \hline
  Noise Source & Noise Contribution\\
  \hline
  G-R Noise at 150 mK & 0.2 eV \\
  Fano Noise in Tantalum & 2.8 eV \\
  Substrate Noise (best) & 12 eV  \\
  Substrate Noise (this device) &  65 eV \\
  \hline
  \hline
\end{tabular}
\caption{A summary of the noise sources present in our resonator.
The noise due to quasiparticle creation and recombination (g-r
noise)\cite{sergeev96} in the aluminum MKID is negligible. The
intrinsic noise of the device from quasiparticle creation statistics
(Fano noise) in tantalum is 2.8 eV.  The dielectric in our
resonators adds phase noise to the measurement\cite{mazin04},
increasing our expected energy width to 65 eV. The excess dielectric
noise displayed by this batch of resonators was significantly worse
than expected from previous measurements due to the use of a
sapphire wafer of poor quality. The best sapphire resonators we have
tested which have the dynamic range to measure 6 keV X-rays would
have given an expected substrate noise contribution of 12 eV.}
\end{table}

The model starts by placing a gaussian distribution of
quasiparticles with a full-width half-maximum (FWHM) of 5~$\mu$m in
a tantalum strip, which is divided into 200 bins. This initial
distribution is propagated forward in time with a time step of
0.1~$\mu$s using the Crank-Nicholson method applied to the diffusion
equation\cite{segall00}.  At each time step, the number of
quasiparticles entering each MKID is recorded.  Perfect
quasiparticle trapping at the interface is assumed. After the
diffusion has been simulated, the quasiparticle pulses are
translated into phase pulses using a simple linear model for MKID
responsivity, $d\theta/dN_{qp} = 1.63\times10^{-7} \alpha Q / V$
radians per quasiparticle, where $\alpha\approx0.07$ is the kinetic
inductance fraction, $Q\approx20,000$ is the resonator quality
factor, and $V$ is the volume of the center strip in
$\mu$m$^3$\cite{mazin04,Gao06b}. These simulated pulses are compared
with the real data, and an iterative routine is used to find the
parameters that best replicate our data. This process is repeated on
ten separate sets of pulses in order to produce error estimates.

Using this model we estimate the diffusion and lifetime parameters
of the 800~$\mu$m long strip since the long length allows the most
accurate determination of the material parameters.  At a temperature
of 150 mK and a microwave readout power at the device of -73 dBm we
measure a tantalum diffusion constant of $13.5 \pm 1.8$ cm$^2$/sec
and a tantalum quasiparticle lifetime of $34.5 \pm 5.7$~$\mu$s.  The
aluminum quasiparticle lifetime is $186 \pm 13$~$\mu$s in the left
MKID and $115 \pm 8.3$~$\mu$s in the right MKID. Similar values are
obtained from other strips.

\begin{figure}
\begin{center}
\includegraphics[width=\columnwidth]{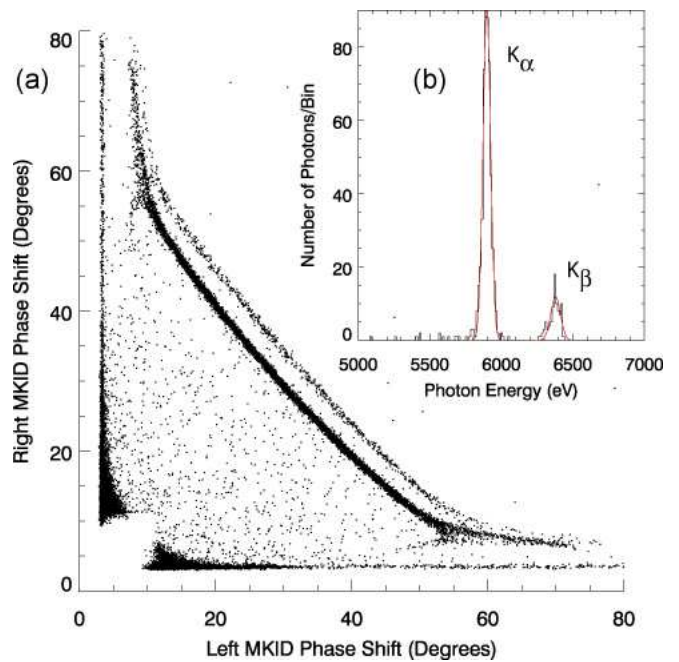}
\end{center}
\caption{The optimally filtered maximum phase pulse height in
degrees observed in an aluminum MKIDs attached to a 200~$\mu$m
tantalum strip is shown in Figure 2(a).  The pulse height in the
left MKID is shown on the x-axis, while the right MKID is shown on
the y-axis. The Mn $K_{\alpha}$ and $K_{\beta}$ lines from the
$^{55}$Fe source are clearly visible. This data is fit to determine
the diffusion length, and this is used to compute the energy
spectrum shown in the inset, Figure 2(b). We calculate a FWHM energy
width $\delta E = 62$ eV at $5.899$ keV when we restrict our data to
all pulses that show greater than 22 degrees of phase shift in both
MKIDs.} \label{fig2}
\end{figure}

These parameters allow us to calculate the tantalum diffusion length
$l_{Ta} = \sqrt{D_{Ta} \tau_{Ta}} = 216 \pm 30$~$\mu$m and relative
responsivity of the MKIDs, which can be used to correct the pulses
for quasiparticle loss in the tantalum strip.  If we define the loss
factor $\beta = l_{strip}/l_{Ta}$ using the values from our model,
the energy of the photons can be calculated from the pulse heights
in each MKID\cite{kraus89}, $P_1$ and $P_2$, using $E = \sqrt{P_1^2
+ P_2^2 + 2 P_1 P_2 \cosh{(\beta)}}$. The inset in Figure 2 shows
the energy histogram derived with this technique for X-ray data
taken at 200 mK. When we consider all pulses with greater than 22
degrees of phase shift in each MKID (the center of the absorber
strip) we obtain a FWHM energy width $\delta E = 62$ eV at $5.899$
keV. This energy width is very close to what we expect from the
observed phase noise in the resonators, which predicts an energy
width of 65 eV. Table I contains a summary of the noise processes in
these detectors.

Improving the magnetic shielding should increase the tantalum
diffusion length substantially\cite{ullom98}, allowing tantalum
absorber strips up to 1 mm long.  In this experiment a cryoperm
magnetic shield was used, but due to a previous change in the
orientation of the experiment it likely did not provide effective
shielding.

The responsivity of an MKID can be increased significantly by using
thinner aluminum films to make the MKID.  Thinner films increase the
kinetic inductance fraction and decrease the volume, so that film
half as thick will have almost four times the responsivity. For a
given maximum photon energy we endeavor to tune the response of the
detector to the largest X-ray to about 90 degrees of phase shift.
Larger phase excursions involve significant heating of the aluminum
film in the MKID, and can make the data taking and analysis more
complex.

If we can reduce our observed noise to the noise we have seen in our
best aluminum on sapphire MKIDs with sufficient dynamic range for 6
keV X-rays, we should be able to get an energy resolution of
$\sim$12 eV, which begins to approach the statistical (Fano) limit
in tantalum (3 eV). Further increases in resolution can be expected
from a more optimal pulse analysis which includes the amplitude
information, and from improvements to MKID design and fabrication
suggested by our ongoing detailed study of their noise properties.

These strips can easily be stacked into a near 100\% fill factor
array, and powerful multiplexing techniques to read out large MKID
arrays have already been demonstrated\cite{mazin06}. These strip
detectors provide a clear path to large format optical/UV and X-ray
focal planes.

The research described in this paper was carried out at the Jet
Propulsion Laboratory, California Institute of Technology, under a
contract with the National Aeronautics and Space Administration.
This work was supported in part by JPL Research and Technology
Development funds and by NASA grant NAG5-5322 to Harrison.  We would
like to thank Rick LeDuc, Jiansong Gao, Dan Prober, and Luigi
Frunzio for useful discussions.

\end{document}